\documentclass{PoS}

\usepackage{epsfig,bm,amsmath}

\newcommand{\onecol}[2]{
        \begin{minipage}[t]{#1}{#2\vfill} \end{minipage}
        }
\def\LALPHA{\hbox{\epsfxsize=2.0 true cm \epsfbox{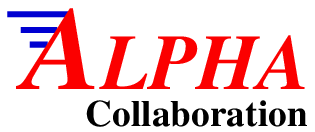}}}

\title{$D_s$ physics from fine lattices}

\ShortTitle{$D_s$ physics from fine lattices}

\author{\LALPHA \hfill
        \onecol{4.0cm}{\vspace{-1.5cm}
          \it DESY 08-117
          \\ SFB/CPP-08-76
          \\ MS-TP-08-24
          \\ HU-EP-08/38
          \vspace{-1.7cm}
        }}
\author{\speaker{Georg von Hippel}, Rainer Sommer\\
        Deutsches Elektronen-Synchrotron DESY,
              Platanenallee~6, 15738~Zeuthen, Germany\\
        Email: \email{georg.von.hippel@desy.de} }
\author{Jochen Heitger\\
        Westf\"alische Wilhelms-Universit\"at M\"unster,
        Institut f\"ur Theoretische Physik,
              Wilhelm-Klemm-Str.~9, 48149~M\"unster, Germany}
\author{Stefan Schaefer\\
        Humboldt-Universit\"at zu Berlin, Institut f\"ur Physik,
              Newtonstr.~15, 12489~Berlin, Germany}
\author{Nazario Tantalo\\
        INFN sezione di Roma ``Tor Vergata'', I-00133~Roma, Italy}

\abstract{We present a preliminary analysis of the charm quark mass
          and the mass and decay constant $f_{D_s}$ of the $D_s$ meson
          obtained from dynamical simulations of $N_f = 2$ Wilson QCD
          on the large and fine lattices simulated by the CLS effort.

          \vspace{2cm}

          Based on CLS configurations
          }

\FullConference{The XXVI International Symposium on Lattice Field Theory \\
                July 14 - 19, 2008\\
                Williamsburg, Virginia, USA}

\begin{document}


\section{Introduction}

Leptonic decays of charmed mesons were not expected to be 
a channel where new physics might be found. However, new, precise
experimental results by CLEO
\cite{Stone:2008gw}
show unexpectedly high rates in the decays $D_{\rm s}\to 
\tau\nu\,,\mu\nu$ compared to estimates from decay constants in the 
quenched approximation. 
The HPQCD collaboration found the effect of (rooted) dynamical
staggered quarks to be significantly smaller than the difference between
experiment and the quenched calculations
\cite{Follana:2007uv}\footnote{Note that a 
relatively new discretization for the charm quarks is used.}.
Is this evidence for new physics
\cite{Dobrescu:2008er},
is it a statistical fluctuation or underestimate of systematics in the 
experiment, or is it a systematic effect unaccounted for by the
errors quoted in
\cite{Follana:2007uv}?
We are aiming at a precise calculation of $f_{D_s}$ as well as other
observables such as the charm quark mass, using the $N_f=2$ CLS 
configurations which reach small lattice spacings, where the charm quark 
mass in lattice units is really small. Here we describe first, encouraging, 
steps. In particular, we find small lattice spacing effects for
O($a$) improved Wilson quarks.


\section{ The CLS coordinated lattice simulations effort }

Coordinated Lattice Simulations (CLS) is a community effort to bring
together the human and computer resources of several teams in Europe
interested in lattice QCD. CLS member teams are located at CERN, in Germany
(Berlin, DESY/Zeuthen, Mainz), Italy (Rome) and Spain (Madrid, Valencia).
All CLS simulations use M. L\"uscher's implementation of the
DD-HMC algorithm
\cite{Luscher:2005rx}
to efficiently simulate $N_f=2$ Wilson QCD with non-perturbative
O($a$) improvement on a variety of computer architectures ranging from
PC clusters to the BlueGene/P at NIC/Forschungszentrum J\"ulich.

Table \ref{tab:ensembles} shows the existing CLS ensembles. For this
initial study of heavy quark physics on these ensembles we will use
the D2, E6 and Q4 ensembles in order to get an idea of the size of
the sea quark mass and lattice spacing effects.

\begin{table}[b]
\parbox{0.5\textwidth}{
\begin{tabular}{lllll}\hline
Id   & Size             & $a$ [fm]  & $\kappa$  & MD $\tau$ \\\hline
D1   & $48\times 24^3$  & $0.08$ & $0.13550$ & $2575$    \\
{\bf D2}   &            &           & $0.13590$ & $2565$    \\
D3   &                  &           & $0.13610$ & $2520$    \\
D4   &                  &           & $0.13620$ & $2505$    \\
D5   &                  &           & $0.13625$ & $2510$    \\\hline
E1   & $64\times 32^3$  & $0.08$    & $0.13550$ & $2672$    \\
E2   &                  &           & $0.13590$ & $2512$    \\
E3   &                  &           & $0.13605$ & $2512$    \\
E4   &                  &           & $0.13610$ & $2497$    \\
E5   &                  &           & $0.13625$ & $2656$    \\
{\bf E6}   &                  &           & $0.13635$ & $4960$    \\
\hline\end{tabular}}
\parbox{0.5\textwidth}{
\begin{tabular}{lllll}\hline
Id   & Size             & $a$ [fm]  & $\kappa$  & MD $\tau$ \\\hline
M1   & $64\times 32^3$  & $0.06$    & $0.13620$ & $4055$    \\
M2   &                  &           & $0.13630$ & $3772$    \\
M3   &                  &           & $0.13640$ & $2980$    \\
M4   &                  &           & $0.13650$ & $3790$    \\
M5   &                  &           & $0.13660$ & $2570$    \\\hline
P1   & $96\times 48^3$  & $0.04$    & $0.13620$ & $1702$    \\
P2   &                  &           & $0.13630$ & started   \\
P3   &                  &           & $0.13640$ & started   \\\hline
{\bf Q4} & $128\times 64^3$ & $0.04$  & $0.13640$ & $1450+$   \\
Q5   &                  &           & $0.13650$ & $1120+$   \\
Q6   &                  &           & $0.136575$ & started  \\
\hline
\end{tabular}}
\caption{\label{tab:ensembles}
  The existing and running CLS ensembles; ensembles used in this
  study are shown in boldface. The molecular dynamics time $\tau$ is given
  in MD units after thermalisation; trajectory length is typically $\tau=0.5$.}
\end{table}

\section{ Setting the scale }

A final determination of the lattice scale, e.g. via
$m_\Omega$, is not yet available for the CLS ensembles. In
\cite{DelDebbio:2006cn}
Del Debbio {\it et al.} determined the scale on the coarsest ($\beta=5.3$)
CLS ensembles to be $a=0.0784(10)$ fm via a combination of $m_K$ and
$m_{K^*}$. We use this as our value for $a$ on the D2 and E6 ensembles,
and run it to $\beta=5.7$ for the Q4 ensemble by means of the scale
$L^*$ defined in
\cite{DellaMorte:2007sb}
via $\bar{g}^2(L^*)=5.5$ in the Schr\"odinger functional scheme. Specifically,
we use the linear fit $\log(L^*/a)=2.3338+1.4025(\beta-5.5)\pm 0.02$.

Since the uncertainty about the scale is an important source of error
at $\beta>5.3$, and the somewhat unphysical determination of the scale
may be considered as a source of an unquantifiable systematic error even
at $\beta=5.3$, a more accurate determination of the scale is certainly a
priority in order to make accurate predictions.


\section{ Measurements and Analysis }

We use 6 time-localized $U(1)$ noise sources per configuration to measure
the correlators $C_{AA}$, $C_{AP}$, $C_{PA}$ and $C_{PP}$, where
\begin{equation}
C_{XY}(x_0) = - a^3 \sum_{{\bf x}} \left<X_{12}(x)Y_{21}(0)\right>,
\end{equation}
$P_{ij}=\bar q_i\gamma_5q_j$ and
$A_{ij}=\bar q_i\gamma_0\gamma_5q_j$, on
      61 configurations of D2,
      28 configurations of E6, and
      31 configurations of Q4,
performing a fully correlated error analysis using the Jackknife method in
each case.

As in
\cite{DelDebbio:2007pz},
we define the effective mass $M_{\rm eff}(x_0)$ via
\begin{equation}
\frac{g(M_{\rm eff}(x_0),x_0-a)}{g(M_{\rm eff}(x_0),x_0)} = \frac{C(x_0-a)}{C(x_0)}
\end{equation}
where
$g(M,x)=e^{-Mx}+e^{-M(T-x)}$.
Effective matrix elements are defined as e.g.
\begin{equation}
G_{PS,\rm eff}(x_0) = \sqrt{\frac{C_{PP}(x_0)M_{\rm eff}(x_0)}{g(M_{\rm eff}(x_0),x_0)}}\;.
\end{equation}
We also define the PCAC quark mass as
\begin{equation}
(m_s+m_c) = m(x_0) = \frac{\frac{1}{2}(\partial_0+\partial^*_0)C_{PA}(x_0)+c_Aa\partial_0\partial^*_0C_{PP}(x_0)}{C_{PP}(x_0)}\,,
\end{equation}
which needs to be renormalized and O($a$) improved as
$m(\mu) = Z_AZ_P^{-1}(\mu)(1+\frac{1}{2}(b_A-b_P)({m}_{q}a)) {m}$.
In terms of these quantities, the (renormalized) pseudoscalar decay constant
is defined as
$F_{PS} = Z_A (1+\frac{1}{2}b_A({m}_{q}a)) \frac{{m} G_{PS}}{M_{PS}^2}$.

We use non-perturbative renormalization wherever possible, in particular
for $c_A$
\cite{Della Morte:2005se},
$Z_A$
\cite{DellaMorte:2008xb},
$Z_P$
\cite{DellaMorte:2005kg},
and $b_A-b_P$
\cite{Heitger:privcomm}.
Perturbative (one-loop) renormalization is used only for $b_A$, where no
non-perturbative results are available. We translate the PCAC masses
into the RGI masses through non-perturbative running in the Schr\"odinger
functional scheme as in
\cite{DellaMorte:2005kg}.
%


\section{Preliminary results}

\subsection{$D_s$ masses and $M_c$ }

\begin{figure}
\begin{center}
\includegraphics[height=0.8\hsize,angle=270,keepaspectratio=]{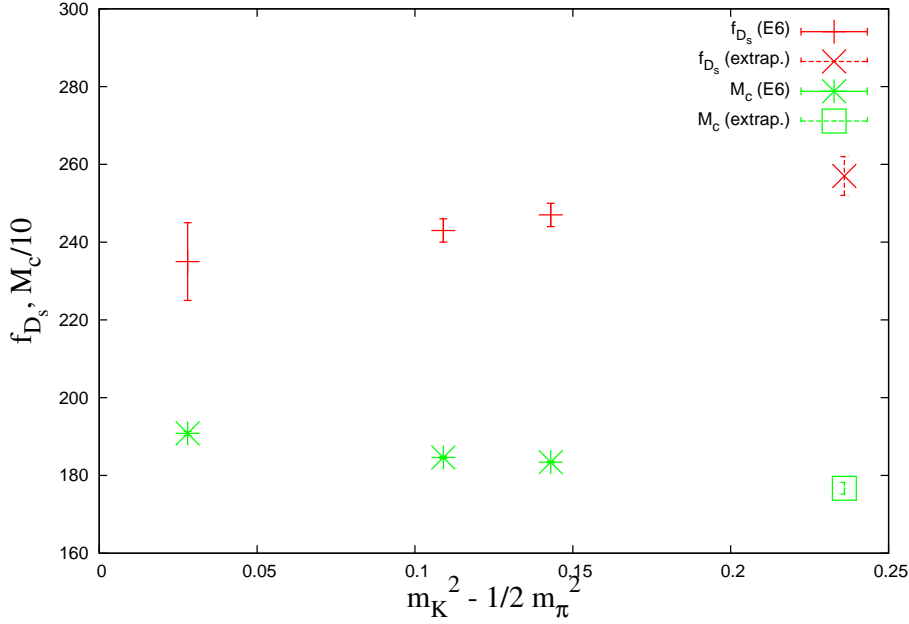}
\end{center}
\caption{\label{fig:fDs_phys}
The quark mass $M_c$ (divided by 10 for scale) and $f_{D_s}$ as a function
of $(m_K^2-\frac{1}{2}m_\pi^2)$ on the E6 ensemble, together with the
linearly extrapolated values at the physical point.}
\end{figure}

Using two heavy quark masses on each ensemble, and two light/strange-quark
masses on E6 and Q4, three light/strange quark masses on D2, and one
additional light quark mass on E6, we measure the masses of all possible
mass combinations of pseudoscalar mesons, as well as the corresponding
PCAC quark masses. From the latter, we extract the RGI mass $M_c$ as
described in 
\cite{DellaMorte:2005kg,Rolf:2002gu}.

To extract a physical value for $M_c$ from these data, we first interpolate
$M_c$ linearly as a function of $M_{D_s}$ to get $M_c$ as a function of the
light and strange quark masses. This we treat as a function of
$(m_K^2-\frac{1}{2}m_\pi^2)$, a $\chi$PT-inspired proxy of the strange quark
mass, and extrapolate to the physical point
$(m_K^2-\frac{1}{2}m_\pi^2)=0.236$ (GeV)$^2$ as shown in fig.
\ref{fig:fDs_phys}.

Our results are $M_c=$ 1694(3)(34) MeV, 1767(15)(35) MeV and 1666(1)(33) MeV
on the D2, E6 and Q4 ensembles, respectively. We note that the lattice
spacing dependence (the 1\% difference between the D2 and Q4 ensembles)
is small, but the sea quark mass dependence
(the 4\% difference between the E6 and D2 ensembles) is noticeable.


\subsection{Lattice spacing effects in $f_{D_s}$}

To get an estimate of what the lattice spacing effects on the decay constant
of the $D_s$ are likely to be, we define an (unphysical) reference point at
$m^{\rm ref}_\pi=m^{\rm ref}_K=618$ MeV, $m_{D}=1968$ MeV, to compare
results obtained at different lattice spacings.
This point is realized directly on the D2 ensemble,
where $L^*f_K^{\rm ref}=0.541(16)(11)$, and $L^*f_D^{\rm ref}=0.805(12)(16)$.
On the Q4 ensemble, we need to interpolate in $m_\pi$ to obtain
$L^*f_K^{\rm ref}=0.578(9)(12)(6)$ and $L^*f_D^{\rm ref}=0.797(9)(16)(9)$.
The errors quoted are from statistics, $L^*$ scale setting and interpolation,
respectively. We find that lattice spacing effects are about $7(5)\%$ in
$f_K^{\rm ref}$, but small in $f_D^{\rm ref}$.

\subsection{$f_{D_s}$ towards the physical point}

To approach the physical point, we take our lightest pion mass, which is
$m_\pi=234(10)(3)$ MeV on the E6 ensemble.

As for the quark mass, we interpolate linearly to $m_{D_s}=1968$ MeV at fixed
$(m_K^2-\frac{1}{2}m_\pi^2)$. A plot of $f_{D_s}$ as a function of
$(m_K^2-\frac{1}{2}m_\pi^2)$ is shown in fig. \ref{fig:fDs_phys}. Extrapolating
to the physical point $(m_K^2-\frac{1}{2}m_\pi^2)=0.236$ (GeV)$^2$, we
obtain our preliminary estimate of $f_{D_s}=257(3)(3)(5)(?)$, where the
question mark denotes unknown systematic errors, including those coming
from the overall scale ambiguity and the quenching of the strange and
charm quarks.

\begin{figure}
\begin{center}
\includegraphics[height=0.8\hsize,angle=270,keepaspectratio=]{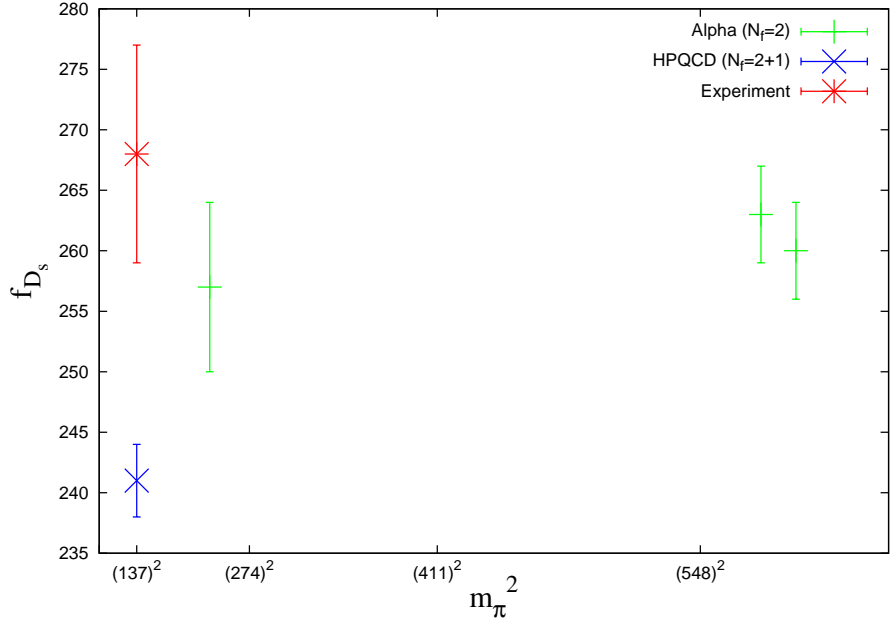}
\end{center}
\caption{\label{fig:summary}
   Summary plot showing the dependence of our results on the pion (sea quark)
   mass, as well as the HPQCD result \cite{Follana:2007uv} and the
   experimental value \cite{Stone:2008gw} for comparison.}
\end{figure}

We summarize our preliminary findings for $f_{D_s}$ in fig. \ref{fig:summary},
which illustrates that cutoff effects are small (at least for heavy pions),
and that the light-quark mass dependence is also small (at least on the coarser
lattice). The chiral and continuum extrapolations therefore seem to be
well possible.

\section{ Summary }

The CLS effort is now simulating very large and fine $N_f=2$ lattices, and
lattice spacings as small as $a=0.04$ fm have become accessible, making 
fully relativistic charm quarks feasible. As simulations are progressing,
lighter sea quarks are also being simulated.

Our preliminary study of the $D_s$ system indicates that cutoff effects are
small and under control, but a more precise scale determination is a priority
in order to eliminate an important source of systematic error.

With better statistics and more sea and valence quark masses to come,
we expect to be able to perform an accurate determination of $f_{D_s}$ in
the near future.


\vskip3ex\textbf{Acknowledgements.}
The authors thank Giulia de Divitiis, Bj\"orn Leder and Roberto Petronzio
for their collaboration in this research project.

We thank NIC/Forschungszentrum J\"ulich for computer time on the BlueGene/P
and BlueGene/L.

This work was supported by the Deutsche Forschungsgemeinschaft in the
SFB/TR 09 and under grant HE 4517/2-1, and by the European community
through EU contract No. MRTN-CT-2006-035482 ``FLAVIAnet''.


\end{document}